\documentclass[fleqn,12pt,twoside]{article}
\usepackage[headings]{espcrc1}

\readRCS
$Id: espcrc1.tex,v 1.2 2004/02/24 11:22:11 spepping Exp $
\usepackage{graphicx}
\usepackage[figuresright]{rotating}

\newcommand{\AmS}{{\protect\the\textfont2
  A\kern-.1667em\lower.5ex\hbox{M}\kern-.125emS}}

\title{Review of Forward Physics at RHIC}

\author{R. Debbe\address[MCSD]{Brookhaven National Laboratory} }
       
\runtitle{Review of Forward Physics at RHIC}
\runauthor{R. Debbe BNL}

\def\Journal#1#2#3#4{{#1}\, {\bf #2}, #3 (#4)}
  
\def\NPA{Nucl.\ Phys.\ A}

\def\PLB{Phys.\ Lett.\ B}
\def\PRL{Phys.\ Rev.\ Lett.}
\def\PRD{Phys.\,Rev.\,D}
\def\PRC{Phys.\,Rev.\,C}

\begin{document}

% typeset front matter
\maketitle

\begin{abstract}
The RHIC high energy collision of species ranging from p+p, p(d)+A to A+A 
provide access to the {\it small-x} component of the hadron wave function. The RHIC program
has brought renewed interest in that subject with its ability to reach values of
the parton momentum fraction smaller than 0.01 with studies of particle production at high rapidity.
Furthermore, the use of heavy nuclei in the p(d)+A collisions facilitates the study of saturation 
effects in the gluonic component of the nuclei because the appropriate scale for that regime grows as 
$A^{\frac{1}{3}}$. We review the experimental results of the RHIC program that have relevance to 
{\it small-x}
emphasizing the physics extracted from d+Au collisions and their comparison to p+p collisions at the same 
energy.
\end{abstract}

\section{INTRODUCTION}

\begin{figure}[htb]
  \resizebox{0.9\columnwidth}{!}
  {\includegraphics[width=15pc]{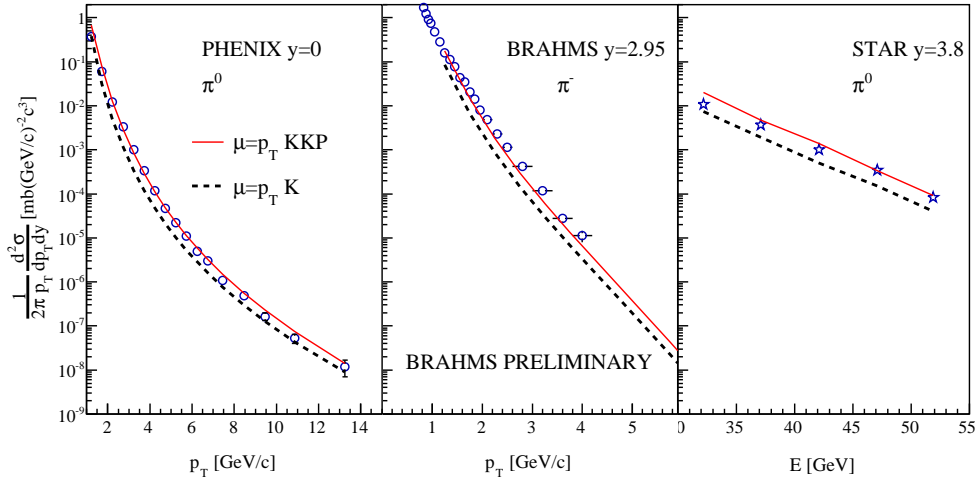}}
  \caption{Invariant cross sections for pion production in p+p collisions at RHIC $\sqrt{s}=200 GeV$
The left-most panel shows the neutral pion distribution measured by PHENIX \cite{PHENIXpi0} at mid-rapidity as
function of transverse momentum. The central panel shows the BRAHMS measurement of negative pions at y=2.95 as 
function of transverse momentum, and the right-most panel shows the neutral pions measured by the STAR collaboration 
\cite{STARpi0} as function of the pion energy. In all three panels the NLO pQCD calculations are shown as smooth curves 
at one scaling parameter ($\mu = p_{T}$) and two sets of fragmentation functions KKP \cite{KKP} and K \cite{Kretzer}. 
All three measurements favor the KKP set of fragmentation functions indicating a strong contribution from gluon-quark 
and gluon-gluon interactions. All errors shown in the figures are statistical.}
  \label{fig:NLO}
\end{figure}

Hadronic interactions in colliders have attained the highest energies up to date. The 
cross-sections of identified pions produced in p+p collisions at RHIC  
($\sqrt{s}= 200 GeV$) are now well described by  Next-to-Leading-Order perturbative 
Quantum Chromo Dynamics (NLO pQCD) calculations at mid and forward rapidities, as 
can be seen in  Fig. \ref{fig:NLO}. This success of QCD allows us to
confidently describe the RHIC measurements above 1-2 GeV/c in $p_{T}$ as related to sufficiently hard 
interactions that can  be calculated with perturbative techniques and partonic degrees of freedom. 
The study of high energy interactions  
 in collider mode opens access to an extended reach  in rapidity space and facilitates the study of 
hadrons wave functions at both extremes of the fractional longitudinal momentum 
$x \sim 0$ and $x \sim 1$. Because the study of particle  
production  at high values of rapidity skews the kinematics of the interaction at the partonic level in 
such a way, that one of interacting partons has a small momentum fraction 
($x \leq 10^{-3}$ around $y=3$ at RHIC), while the second one  carries most of the beam momentum.

\section{BRAHMS}

\begin{figure}[htb]
  \resizebox{0.9\columnwidth}{!}
  {\includegraphics[width=15pc]{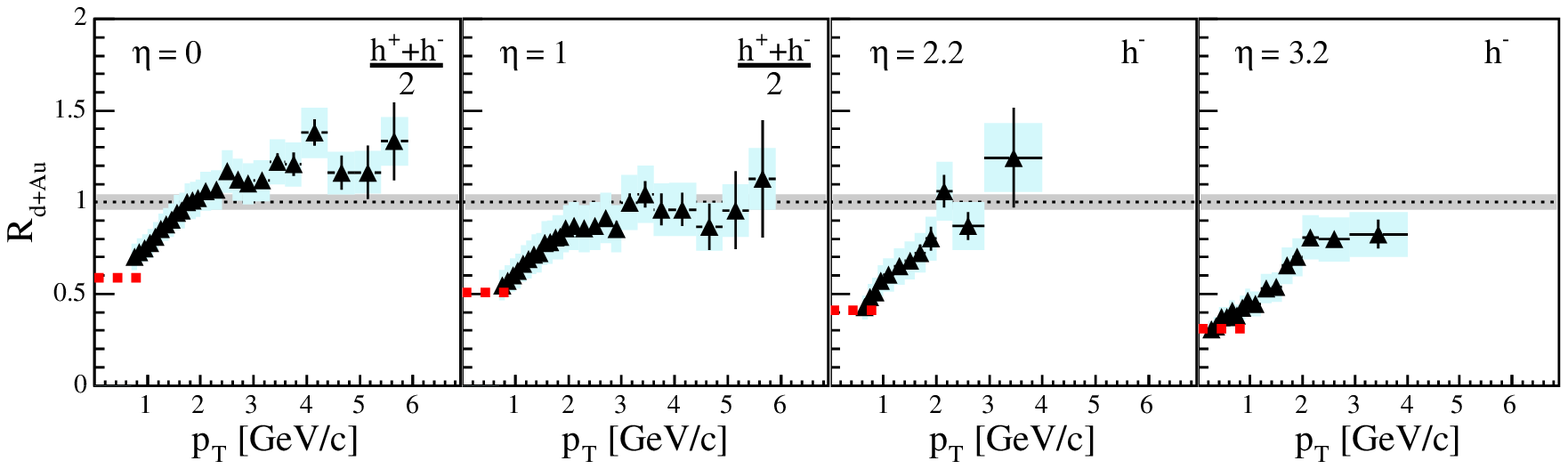}}
  \caption{Nuclear modification factor extracted from minimum biased sample of charged particles at $\eta=0$ 
and $\eta=1$. The same factor, but this time for negative hadrons is shown in the two right-most panels for 
$\eta=2.2$ and $\eta=3.2$. The mean number of collisions 
used to make this factor is $7.2 \pm 0.3$. Statistical errors are shown with vertical lines and systematic errors 
with gray boxes.}
  \label{fig:RdA}
\end{figure}

The Forward Spectrometer (FS) of the BRHAMS Collaboration is particularly well suited to study charged particle 
production at high rapidity. The complement of tracking, Cerenkov's ring imaging and time-of-flight techniques
have been used to extract new physics from the comparison of invariant yields of charged particles in d+Au and p+p 
collisions at the same energy. Such comparison is shown in Fig. \ref{fig:RdA} with the so called nuclear modification 
factor $R_{dAu}$ \cite{PRL93dA}.
The changes in magnitude and shape of the $R_{dAu}$ factor seen in this figure as rapidity values 
grow from mid-rapidity ($\eta=0$) on the left-most panel all the way to the highest value ($\eta=3.2$) shown
on the right-most panel, are consistent with the onset of the Color Glass Condensate (CGC) \cite{CGC} at RHIC.

\begin{figure}[!htb]
  \resizebox{0.9\columnwidth}{!}
  {\includegraphics[width=15pc]{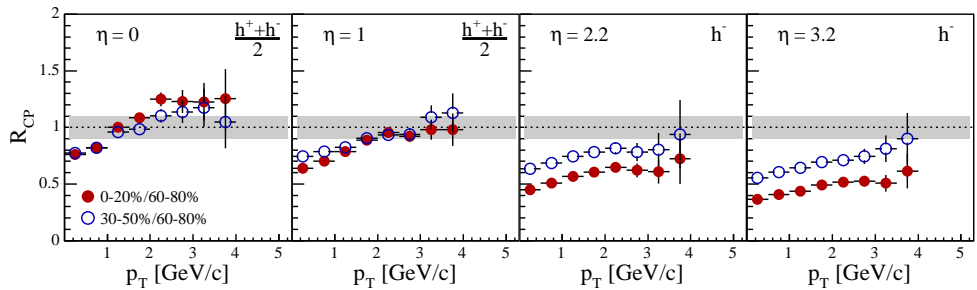}}
  \caption{Nuclear modification factor extracted from the comparison of d+Au data samples with different 
centralities at the same pseudo-rapidities as Fig. \ref{fig:RdA}.}
  \label{fig:Rcp}
\end{figure}

 The CGC 
is expected to appear as a form of high density gluonic matter as the momentum fraction of the partons tends to zero.
 In a CGC based description of d+Au collisions, the partons in the Au wave function tend to populate states
with transverse momentum values around the saturation scale $Q_s$ and the valence quarks from the deuteron
beam multiple scatter on them producing the enhancement seen in Fig. \ref{fig:RdA} at $\eta=0$ . 
Away from $\eta=0$, the Au wave function is further modified by gluon emission (the probability of gluon emission per
unit of rapidity is proportional to $\alpha_s$ the strong coupling constant).  Gluons with small values of $x$ are 
delocalized and the gluon fields extend into other nucleons. Besides that, gluons have as well a strong
tendency to fuse and the evolution of the wave function with rapidity ends up as a compromise between  gluon emission
and gluon fusion that tends toward a limiting value. The comparison between d+Au and p+p yields shown with the $R_{dAU}$ 
ratio will show an stronger suppression as rapidity grows because the numerator would be reaching the above mentioned 
limiting value, while the denominator is still growing with rapidity because that system is considered dilute 
and is not affected by the small contribution from gluon fusion \cite{Wiedemann,KKT}. 
Such behavior can be seen in the measured rapidity dependence of the $R_{dAu}$ factor displayed in
 Fig. \ref{fig:RdA}. The effects of quantum evolution are more pronounced in central
events. Central events have the highest number of interacting nucleons that, as mentioned above, share their {\it small-x}
gluon fields increasing the chances for gluon fusion. These systems with higher number of gluons would reach the limiting 
value of their evolution with rapidity much earlier than the nuclei interacting with bigger impact parameters. 
Fig. \ref{fig:Rcp} $R_{cp}$ has been constructed with three data samples: central events with centralities in 
the 0-20\% range, mid-central events in the 
range 30-50\% and the peripheral events used as reference with centralities in the 60-80\% range. At mid-rapidity the
central $R_{cp}$ points are systematicaly above the semi-central ones, but at the highest rapidity ($\eta=3.2$), the 
trend is reversed, and this time the central events have a very much suppresed value of $R_{cp}$ compared to the
semi-central sample. 

\section{PHOBOS}

The PHOBOS Collaboration has presented similar results as the ones shown in the previous section. They measure the 
yields of charged hadrons with their two-arm magnetic spectrometer equipped with high resolution silicon based tracker.
They have extracted the nuclear modification factor from a centrality averaged data sample of charged hadrons in three 
pseudo-rapidity bins, $0.2 < \eta < 0.6$, $0.6 < \eta < 1.0$ and $ 1.0 < \eta < 1.4$. They use charged hadron spectra 
measured by the UA1 Collaboration \cite{UA1} in $p + \bar{p}$ at the same energy, and a suitable correction is
applied to account for the difference in acceptance between the two experiments. 

\begin{figure}[htb]
\begin{center}
  \resizebox{0.6\columnwidth}{!}
  {\includegraphics[width=15pc]{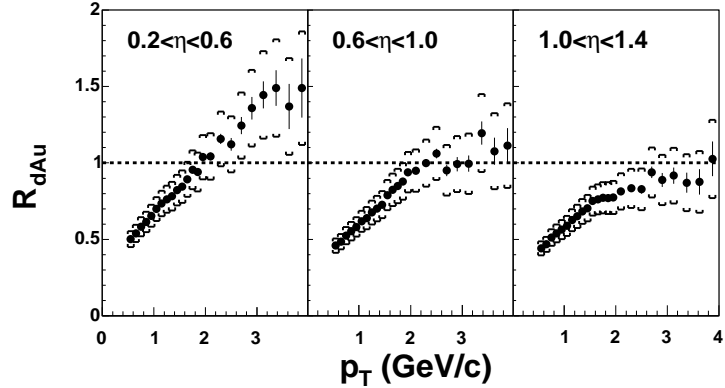}}
  \caption{Nuclear Modification factor $R_{dAu}$ for charged hadrons measured in the PHOBOS two-arm spectrometer as function of the transverse momentum in three pseudo-rapidity bins. The brackets show the systematic errors.}
  \label{fig:Phobos_RdAu}
\end{center}
\end{figure}

Figure \ref{fig:Phobos_RdAu} shows the nuclear modification factor as a function of the 
transverse momentum, The mean number of collisions used to construct this factor is equal to $9.5 \pm 0.8$(systematic). 
The remarkable change in the shape of the $R_{dAu}$ factor as 
the pseudo-rapidity changes by only half a unit has been highlighted as an indication of the fast onset
of the effects of quantum evolution on an already dense Au wave function.

\section{PHENIX}

The PHENIX collaboration has measured the yield of charged hadrons with their muon spectrometers 
covering the Au fragmentation region (backward) $-2.2 < \eta < -1.2$, and the deuteron fragmentation
$1.2 < \eta < 2.2$ (forward). Although these spectrometers are optimized to detect muons, they can also be used 
to detect charged particles that penetrate the spectrometer up to its last active
stages. These penetrating particles are designated as "punch-through hadrons" (PTH). 
A second method detects the muons from 
hadrons that decay before the first layer of absorber in the spectrometer, these particles are referred as 
"hadron decay muons" (HDM) \cite{PHENIX_PRL94dA}.

\begin{figure}[htb]
\begin{center}
  \resizebox{0.6\columnwidth}{!}
  {\includegraphics[width=15pc]{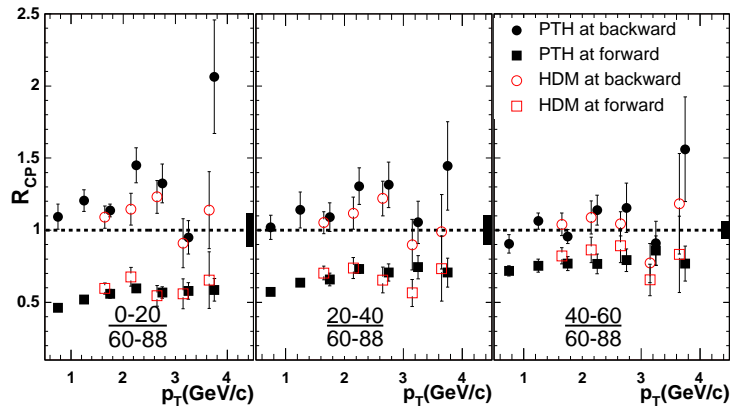}}
  \caption{Nuclear modification factor $R_{cp}$ as function of $p_{T}$ in the forward (squares) and backward 
rapidity (circles). These factors are extracted from four data samples with different 
centralities. The centrality of the event is measured with the backward BBC detectors.}
  \label{fig:PHENIX_Rcp}
\end{center}
\end{figure}

Figure \ref{fig:PHENIX_Rcp} shows, first of all, the consistent results obtained with the PTH and 
HDM techniques. The measurement performed on the Au fragmentation side (backward) shows a weak 
enhancement that is most pronounced in the most central sample ($\frac{0-20\%}{60-88\%}$). A strong suppression 
on the deuteron fragmentation side is seen in the central event sample. Such suppression is reduced
as the centrality selection changes in the ($\frac{40-60\%}{60-88\%}$) ratio. The PHENIX forward factors are consistent 
with the ones measured by BRAHMS and PHOBOS. The backward 
measurements probe intermediate values of the momentum fraction $x$ in the Au wave function,  
the hint of an enhancement remains unexplained and it could be a combination of different effects \cite{PHENIX_PRL94dA}.

\begin{figure}[htb]
\begin{center}
  \resizebox{0.6\columnwidth}{!}
  {\includegraphics[width=15pc]{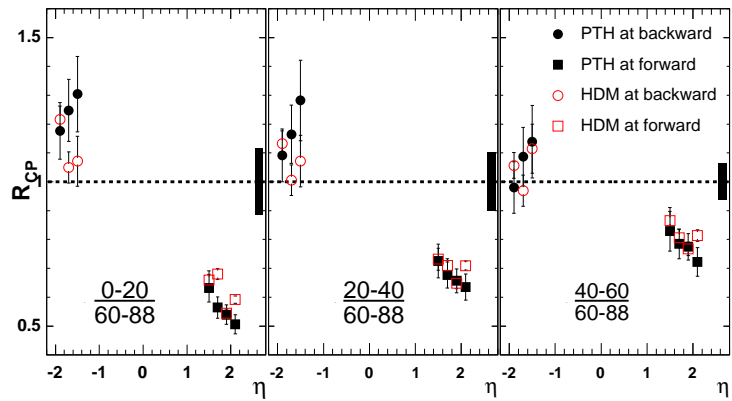}}
  \caption{The integrated Nuclear modification factor $R_{cp}$ as function of pseudo-rapidity measured in both the Au side (backward) and the deuteron side (forward) extracted from four different
centrality data samples.}
  \label{fig:PHENIX_RcpVsEta}
\end{center}
\end{figure}

Figure \ref{fig:PHENIX_RcpVsEta} shows the integrated $R_{cp}$ distributions  
 over the intervall $ 1.5 < p_{T} < 4.0$ GeV/c at different values of pseudo-rapidity
within the acceptance of both muon spectrometers. The main feature on this figure is the strong suppression
in the forward side (deuteron fragmentation) for the most central events, as well as a  clear
enhancement in the Au fragmentation side for the same data sample.

\section{STAR}

The STAR collaboration has measured the yield of high energy neutral pions ($25 < E_{\pi} < 55$ GeV)
 at high rapidity ($3.0 \leq \eta \leq 4.2$) in p+p and d+Au collisions at 200 GeV with their
forward $\pi^{0}$ detector (FPD) \cite{STAR_RdA}. Figure 7 shows the nuclear 
modification factor $R_{dAu}$ for neutral pions (filled circles) at $<\eta>=4$, 
the highest rapidity studied at RHIC so far. The figure also shows the BRAHMS factors
extracted at $\eta = 2.2$ (open circles), and $\eta=3.2$ (open squares). Both BRAHMS and STAR data
were extracted without conditions on centrality. The suppression seen in the STAR $\pi^{0}$ factor
is stronger than the one measured with negative hadrons and both measurements are consistent if
the isospin effects in p+p are taken into account as well as the higher rapidity value of the STAR
measurement. The inset shows the STAR $R_{dAu}$ factor together with ratios built from NLO pQCD calculations 
with the KKP and K sets of fragmentation functions and a third calculation that includes
coherent multiple scattering in the Au target \cite{VitevQiu}. The suppression found in the data is clearly much 
stronger than the calculations. For more details please refer to \cite{STAR_RdA}. 

\begin{figure}[htb]

\begin{center}
 \resizebox{0.4\columnwidth}{!}
  {\includegraphics{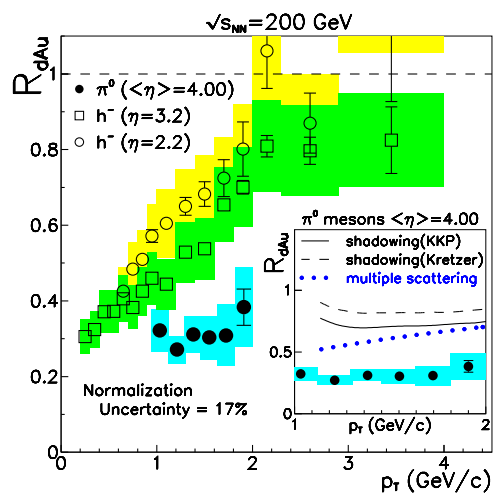}}
\end{center}

  \label{fig:StarRdA}

  \caption{STAR nuclear modification factor ($R_{\rm dAu}$) for minimum-bias 
neutral pion production in d+Au collisions versus transverse momentum ($p_T$), shown with filled circles.
The BRAHMS factors extracted for negative hadrons 
 smaller at $\eta=2.2$ and $\eta=3.2$ are also shown with open circles and boxes respectively \cite{PRL93dA}.
The error bars are statistical, and the shaded boxes show the 
point-to-point systematic errors.
(Inset) The STAR $R_{\rm dAu}$ compared to the ratio of NLO pQCD calculations and \cite{VitevQiu}. 
}
\end{figure}

\begin{figure}[htb]

\begin{center}
  \resizebox{0.4\columnwidth}{!}
  {\includegraphics[scale=0.5]{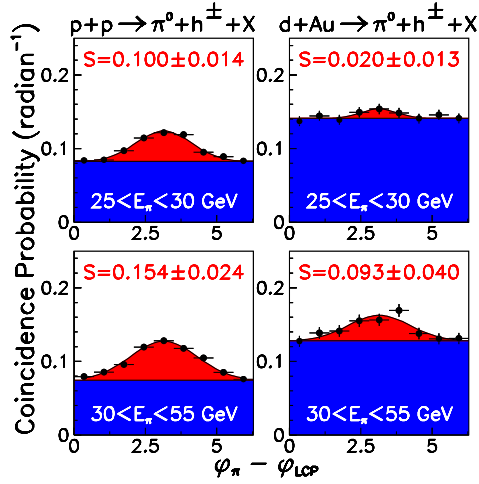}}
\end{center}

  \caption{Azimuthal correlations centered around $\Delta \phi = \pi$ between a forward high energy neutral pion 
and charged hadrons detected around mid-rapidity in the STAR TPC. The left column shows the correlation obtained from
p+p events at two pion energy bins: $25.<E_{\pi} < 30$ GeV (top)  and $30.<E_{\pi} < 55$ GeV (bottom). The right 
column shows the same correlation extracted from d+Au collisions. The errors are 
statistical. For further details see \cite{STAR_RdA}.}
  \label{fig:StarAzimuth}
\end{figure}

STAR can correlate high energy neutral pions at high rapidity and charged hadrons detected in the 
Time-Projection-Chamber (TPC). The difference in azimuth angle 
between the forward neutral pion, and charged particles
detected in the TPC ($|\eta|<0.75$) with transverse momentum greater than 0.5 GeV/c is shown in 
Fig. \ref{fig:StarAzimuth}. Azimuth angle correlations measured in p+p collisions are shown in the left column of 
the figure and the ones extracted from d+Au events are shown on the right column. The effects of gluon emission 
between the parton that generates the forward pion and the one related to the charged particle detected near mid-rapidity 
would destroy the back-to-back correlation and would appear as the production 
of mono-jets \cite{kharzeevmonojet}. Such effect on the azimuthal angle correlation is visible in the low pion 
energy bin, but is not present in the second, higher energy bin. This result is tantalizing and has been presented 
as another indication of the onset of coherence in the CGC. The PHENIX collaboration has extracted similar correlations 
and they found no indication of de-correlation of back-to-back jets. The PHENIX measurement was performed with hadrons 
separated by a smaller rapidity gap than the STAR measurement \cite{PHENIX_Monojet}. Further studies of this nature are 
expected to be performed in near future RHIC runs.  

Other theoretical approaches have been offered to explain the suppression at high rapidity in d+Au collisions. 
Such is the case of a description of d+A collision as coherent scattering equivalent to a shift in x \cite{VitevQiu}.
Perturbative QCD calculations with a standard shadowing parametrization added to the parton distribution functions 
 \cite{shadowing} are also available. Other calculations emphasize the effects of
parton recombination \cite{recombination}, and finally, one of the latest explanations to the high rapidity suppression 
focuses on energy conservation arguments that may be applicable for deuteron valence quarks with high $x$ 
values \cite{factorization}

\section{FUTURE PROSPECTS}

Two of the  RHIC experiments (BRAHMS and PHOBOS) have by now ended their data taking phase, and the STAR
and PHENIX collaborations are well into a detector upgrade program aimed to increase 
their rapidity coverage, and to enhance their ability to detect rare processes connected to heavier quarks. These 
upgrades are scheduled to come on-line as RHIC evolves into its second phase 
when it will provide collisions at higher luminosity.
The STAR collaboration is working on a bigger Lead-glass array called the Forward Meson Spectrometer (FMS) placed 
at the same location of
the existing FPD on the detuteron fragmentation side. This detector will have a rapidity coverage of 
$2.5 < \eta < 4.0$. The FMS will be
used to continue the study of correlations between high energy mesons and photons and charged particles detected 
with the TPC in search for the onset of saturation in possible "macroscopic gluons fields" \cite{LesBland}.

The PHENIX collaboration is adding precision tracking before the first muon absorber with the Forward Silicon Tracker. 
They will also replace the "nose-cone"
hadronic absorbers in the muon arms with a highly segmented Si-W calorimeters dubbed Nose Cone Calorimeter (NCC). 
This addition to the PHENIX  
detector will extend its pseudo-rapidity coverage up to a value of 3 to allow for precision measurements of 
direct photons, 
$\pi^{0}$s and di-electrons in A+A, p(d)+A, and polarized p+p collisions, in particular it will be able to identify pions 
with momentum as high as 30 GeV/c. Together with the Forward Silicon Tracker, the NCC can be used to measure
 D and B meson productions via their semi-leptonic decays.

\section{SUMMARY}

The study of the RHIC d+Au collisions, together with results extracted from the several p+p 
RHIC runs, have renewed the interest in the {\it small-x} physics of hadronic systems. Together with the lower than 
expected multiplicities detected in A+A collisions at top RHIC energies, these results indicate the onset of a 
saturated regime that exhibits coherence. Further studies
of d+Au collisions at high rapidity are planed for the near future by the two big RHIC experiments. 
These measurements will include a search of mono-jets as well and a shift to probes
that have a more direct connection to the gluon component of the Au nuclei such as direct 
virtual and real photon production as well as open charm detection at high rapidity .

\end{document}